\documentstyle[preprint,aps]{revtex}
\tightenlines
\begin{document}

\title{Simple Scheme for Efficient Linear Optics Quantum
Gates}
\author{T.C.Ralph, A.G.White, W.J.Munro and G.J.Milburn}
\address{Centre for Quantum Computer Technology,
\\ University of Queensland, QLD 4072, Australia
\\Fax: +61 7 3365 1242  Telephone: +61 7 3365 3412 \\
email: ralph@physics.uq.edu.au\\}
\date{23rd March 2001}
\maketitle

\begin{abstract}
We describe the construction of a conditional quantum control-not (CNOT)
gate from linear optical elements following the program of Knill, Laflamme
and Milburn [Nature {\bf 409}, 46 (2001)]. We show that the basic
operation of this gate can be tested using current technology.
We then simplify the scheme
significantly.
\end{abstract}
\narrowtext

\vspace{5mm}

\section{Introduction}

Optics would seem to be a strong contender for realizing quantum
computation circuits. Photons are
easily manipulated and, as the electro-magnetic
environment at optical frequencies can be regarded as vacuum,
are relatively decoherence free. Indeed one of the earliest proposals
\cite{mil88} for
implementing quantum computation was based on encoding each qubit in
two optical modes, each containing exactly one photon. Unfortunately,
2 qubit gates require strong interactions between single photons. Such
interactions would require massive, reversible non-linearities well beyond
those
presently available.

Recently Knill , Laflamme and Milburn (KLM) found a way to circumvent this
problem and implement efficient quantum computation using only passive
linear optics, photodetectors, and single photon sources \cite{kni00}.
This efficient
linear optical quantum computing (ELOQC) is distinct from other linear
optical schemes \cite{kwiat} which are not efficiently scalable.

Although containing only linear elements, the optical networks
described by KLM are complex and would present major stability and
mode matching problems in their construction. There is thus
considerable interest in finding the simplest physical implementations
of the KLM techniques. In this manuscript we investigate this problem and
find a major simplification of the original proposal.

We begin by reviewing the technique via which non-deterministic gates 
can be used to implement an efficiently scalable system and in 
Section 3 
the physics of a basic non-deterministic gate, the NS 
gate, is discussed.
In Section 4 we describe the construction of a non-deterministic
quantum CNOT gate using two NS gates. Full scalability of this
gate requires high efficiency, 0, 1, 2 photon discriminating photon
counters. Such detectors presently only exist in prototype form
\cite{rock}. However, in Section 5 we show that the basic
operation of this circuit can be tested with current detector
technology. We then describe the simplified gate. 

A non-deterministic
CNOT gate with a simple linear architecture, but requiring triggered
entangled sources as a resource, has been suggested recently \cite{imo}.
In contrast our scheme requires only separable input states. Also 
recently proposed is a linear optical scheme for the probabilistic purification 
of non-maximal polarization entangled states \cite{pan}. Although the 
linear elements play the role of CNOT gates in this protocol, they do 
not exhibit the full CNOT logic of the gates described here.

\section{Gate Operation via Teleportation}

Arbitrary quantum gate operations can be implemented if one has the
ability to implement arbitrary single qubit rotations and two qubit CNOT
gates. Single qubit operations can easily be implemented with single
photons and a non-deterministic CNOT gate is described in this
manuscript. However a cascaded sequence of
such non-deterministic gates would be useless for quantum computation because
the probability of many gates working in sequence decreases
exponentially. This
problem may be avoided by using a teleportation protocol \cite{benn}
to implement quantum gates. The
idea that teleportation can be used for universal quantum computation was
first proposed by Gottesman and Chuang \cite{gott}.

A teleportation circuit is represented in Fig.1(a). A qubit in an unknown
state $|\alpha \rangle$ is teleported by making a joint Bell
measurement ($B$)
of it and one half of a Bell pair $|\Phi \rangle$. Depending on the
result of the measurements, $\sigma_{x}$ and $\sigma_{z}$ manipulations
are made on the other half of the Bell pair resulting in accurate
reconstruction of the unknown state. A key issue is that the Bell
pair plays the role of a resource in the protocol. That is, it can be prepared
``off-line'' and then used when necessary to teleport the qubit.
Now consider the quantum
circuit shown in Fig.1(b). Two unknown qubits are individually
teleported and then a CNOT gate is implemented. Obviously, but
not very usefully, the result is CNOT operation between the input and
output qubits. However, as shown in Ref.\cite{gott}, the commutation
relations between CNOT and $\sigma_{x}$ and $\sigma_{z}$ are quite
simple, such that the circuits of Fig.1(b) and 1(c) are in fact
equivalent. But in the circuit of Fig.1(c) the problem of implementing a
CNOT gate has been reduced to
that of producing the required entanglement resource. The
entanglement resource required could be produced from separable input 
states
using three CNOT gates: one each to produce the Bell pairs plus the 
one shown in Fig.1(c). But the point is that these need not be
deterministic gates. Non-deterministic CNOT gates could be used in
a trial and error manner to build up the necessary resource off-line.
It could then be used when required to
implement the gate.

A remaining issue is the performance of the Bell measurements
required in the teleportation protocol. These cannot be performed
exactly with linear optics. KLM
showed that by using the appropriate entangled resource the teleportation step
can be made near deterministic. The near deterministic teleportation protocol
requires only linear optics, photon counting and fast feedforward, 
albeit with a significant resource overhead.
Alternatively, progress has recently been made towards implementing
Bell measurements using non-linear optics \cite{shih}.

\section{The NS Gate}

The basic element in the construction of our non-deterministic CNOT
gate is the nonlinear sign-shift (NS) gate \cite{kni00}. This is
a non-deterministic gate the operation of which is
conditioned on the detection of an auxiliary photon. When successful
the gate implements the following transformation on signal state
$|\psi \rangle$
\begin{equation}
|\psi\rangle=\alpha |0\rangle+\beta |1\rangle+
\gamma |2\rangle \to |\psi'\rangle=
0.5(\alpha |0\rangle+\beta |1\rangle- \gamma |2\rangle)
\label{NS}
\end{equation}
where the lack of normalization of the transformed state reflects the
fact that the gate has a probability of success of $0.25=(0.5)^{2}$.

Fig.2 shows a realization of this gate. Two ancilla modes
are required. A single photon is injected into one of the ancilla and the
other is unoccupied. The first, second and third beamsplitters have
intensity reflectivities $\eta_{1}$, $\eta_{2}$ and $\eta_{3}$
respectively. The beamsplitters are phase asymmetric:
transmission from either side and reflection off the ``black'' surface of
these
beamsplitters results in no phase change, whilst reflection off the
``grey'' surface results in a sign change. 
When a single photon is counted at the ``1'' ancilla output and
no photon is counted at the ``0'' ancilla output (as indicated in the figure)
the transformation of Eq.\ref{NS} is implemented if a suitable choice of
beamsplitter reflectivities is made. Let us see how this works.

Suppose first that the signal mode is in the vacuum state, i.e.
$|\psi\rangle=|0\rangle$. The probability amplitude, $C$, for the
ancilla  photon to
appear at the ``1'' output port is given by
\begin{equation}
C=\sqrt{\eta_{1} \eta_{2} \eta_{3}}+\sqrt{(1-\eta_{1})(1-\eta_{3})}
\label{NS1}
\end{equation}
Now suppose the input is a single
photon state, i.e. $|\psi\rangle=|1\rangle$. If a photon arrives at the ``1''
output port and no photon arrives at the ``0'' port then a single
photon must have exited the signal output. We wish the probability
amplitude for this event to also be $C$. This means
\begin{eqnarray}
C & = & \sqrt{\eta_{1} \eta_{3}} (1-\eta_{2})-(\sqrt{\eta_{1} \eta_{2}
\eta_{3}}+
\sqrt{(1-\eta_{1})(1-\eta_{3})})\sqrt{\eta_{2}} \nonumber\\
 & = & \sqrt{\eta_{1} \eta_{3}} (1-\eta_{2})-C \sqrt{\eta_{2}}
 \end{eqnarray}
 and thus
\begin{equation}
C={{\sqrt{\eta_{1} \eta_{3}} (1-\eta_{2})}\over{1+\sqrt{\eta_{2}}}}
\label{NS2}
 \end{equation}
 Finally we consider the situation of a two photon input, i.e.
 $|\psi\rangle=|2\rangle$. If a single photon arrives at the ``1'' port
 and no photon arrives at the ``0'' port then two photons must have
 exited at the signal output. To obtain the sign change of Eq.\ref{NS}
 we require the probability amplitude for this event to be $-C$. This
 means
\begin{eqnarray}
-C & = & -\sqrt{\eta_{1} \eta_{3} \eta_{2}}(1-\eta_{2})-
\sqrt{\eta_{2}}(\sqrt{\eta_{1} \eta_{3}} (1-\eta_{2})-
(\sqrt{\eta_{1} \eta_{2} \eta_{3}}+
\sqrt{(1-\eta_{1})(1-\eta_{3})})\sqrt{\eta_{2}}) \nonumber\\
 & = & \eta_{2} C-2 \sqrt{\eta_{1} \eta_{2} \eta_{3}}(1-\eta_{2})
 \label{NS3}
 \end{eqnarray}
Substituting Eq.\ref{NS2} into Eq.\ref{NS3} gives the result
\begin{equation}
\eta_{2}=(\sqrt{2}-1)^{2}
\end{equation}
Substituting back into Eq.\ref{NS2} and Eq.\ref{NS1} we can solve for
$\eta_{1}$,
$\eta_{3}$ and $C$. The maximum value for $C$ is achieved when
\begin{equation}
\eta_{1}=\eta_{3}={{1}\over{(4-2\sqrt{2})}}
\end{equation}
and is
\begin{equation}
C=0.5
\end{equation}
Thus the transformation of Eq.\ref{NS} is implemented whenever a single photon
is recorded at port ``1'' and no photon is found at port ``0''. On
average this will occur 25\% of the time since $|C|^{2}=0.25$.

\section{The CNOT Gate}

A conditional CNOT gate can now be implemented using two NS
gates. The layout for doing this is shown schematically in Fig. 3.
We employ dual rail logic such that the ``control in'' qubit is
represented by the two bosonic mode operators $c_{H}$ and $c_{V}$. A
single photon occupation of $c_{H}$ with $c_{V}$ in a vacuum state
will be our logical 0, which we will write $|H\rangle$ (to avoid confusion
with the vacuum state). Whilst a
single photon occupation of $c_{V}$ with $c_{H}$ in a vacuum state
will be our logical 1, which we will write $|V\rangle$. Of course
superposition states can also be formed. Similarly the ``target in''
is represented by the bosonic mode operators $t_{H}$ and $t_{V}$ with
the same interpretations as for the control. The beamsplitters, $B1$,
$B2$, $B3$ and $B4$ are all 50:50.

The four modes $c_{H}$, $c_{V}$, $t_{H}$ and $t_{V}$ are all the same 
polarization. The use of the ``H'', ``V'' nomenclature alludes to the standard 
situation in which the two modes of the dual rail logic are orthogonal 
polarization modes. Conversion of a polarization qubit into the 
spatial encoding used to implement the CNOT gate can be achieved 
experimentally by passing the the photon through a polarizing 
beamsplitter, to spatially separate the modes, and then using a 
half-wave plate to rotate one of the modes into the same polarization 
as the other. After the gate, the reverse process can be used to 
return the encoding to polarization.

The layout of Fig. 4 contains two nested, 
balanced Mach-Zehnder interferometers. The
target modes are combined and then re-separated forming the
``T''interferometer.
One arm of the T interferometer and the $c_{V}$ mode
of the control are combined to form another interferometer, the ``C''
interferometer. NS gates
are placed in both arms of the C interferometer. The essential feature 
of the system is that if the control photon is in the $c_{H}$ mode 
then there is never more that one photon in the C interferometer, 
so the NS gates do not produce a change, the T interferometer remains 
balanced and the target qubits exit in the same spatial modes in 
which they entered. On the other hand if the control photon is in 
mode $c_{V}$ then there is a two photon component in the C 
interferometer which suffers a sign change due to the NS gates. This 
leads to a sign change in one arm of the T interferometer and the 
target qubit exits from the opposite mode from which it entered.

Let us consider the systems operation in more detail: 
If the control is in a logical 0 then the
mode $c_{V}$ will be in a vacuum state. Consider the line labeled $x$
in Fig.3 lying just before the NS  gates. The state of the system at
this point is given by
\begin{equation}
|\psi \rangle_{x}={{1}\over{\sqrt{2}}}|1001 \rangle \pm {{1}\over{2}}(|1100
\rangle - |1010 \rangle )
\end{equation}
where the left to right ordering is equivalent to the top to bottom 
ordering in Fig.3.
The $+$ occurs when the target input state is $|H \rangle$, the $-$ occurs
when
the target input state is $|V \rangle$. Now consider the state of the system
directly
after the NS gates operate on the middle two modes (indicated by the
line $y$ in Fig.3). Substituting from Eq.\ref{NS} we find $|\psi
\rangle_{y}=0.25 |\psi \rangle_{x}$. That is the
gates do not effect the states in the arms of the C
interferometer (conditional on the detection of photons at the ``1''
ports of the NS gates). As both interferometers are balanced they will
just return the same outputs as they had inputs. Thus $c_{Vo}$ will be
a vacuum mode, and if the target input photon was in $t_{H}$, it will
emerge in $t_{Ho}$; or if it was in $t_{V}$, it will emerge in $t_{Vo}$.
In other words the control and target qubits will remain in the same
states.

On the other hand if the control is in a logical 1, then the $c_{V}$
mode will contain one photon. The state at $x$ is now
\begin{equation}
|\psi \rangle_{x}={{1}\over{2}}(|0101 \rangle + |0011 \rangle \pm(|0200
\rangle - |0020 \rangle) )
\end{equation}
The two photon
amplitudes suffer sign changes (conditional on the detection of
photons at the ``1'' ports of the NS gates) such that the state at
$y$, after the NS gates, is now
\begin{equation}
|\psi \rangle_{y}=0.25 ({{1}\over{2}}(|0101 \rangle + |0011 \rangle
\mp (|0200
\rangle - |0020 \rangle )))
\end{equation}
This leads to a sign change in the returning beam of the T
interferometer which in turn results in a swap between the inputs
and outputs of the T interferometer. Thus if the target input photon
was in $t_{H}$ it will
emerge in $t_{Vo}$ or if it was in $t_{V}$ it will emerge in $t_{Ho}$.
The control output, $c_{Vo}$ also suffers a sign change, but this
does not change its logical status.
In other words the control is unchanged but the target qubit will
change states.

The truth table of the device is thus
\begin{eqnarray}
    |H\rangle_{c} |H\rangle_{t} & \to & |H\rangle_{c} |H\rangle_{t},
    \; \; \; \;
    |H\rangle_{c} |V\rangle_{t}  \to  |H\rangle_{c} |V\rangle_{t} \nonumber\\
    |V\rangle_{c} |H\rangle_{t} & \to & |V\rangle_{c} |V\rangle_{t},
    \; \; \; \;
    |V\rangle_{c} |V\rangle_{t}  \to  |V\rangle_{c} |H\rangle_{t}
    \label{logic}
    \end{eqnarray}
Which is CNOT logic.

It is useful to also look at this arrangement in the Heisenberg picture.
Referring again to Fig.3
our input modes are $c_{H}$ and
$c_{V}$ for the control, $t_{H}$ and $t_{V}$ for the target, and the
ancilla modes $a_{1}$, $a_{2}$, $v_{1}$ and $v_{2}$. The initial
state of $c_{i}$, $t_{j}$, $a_{1}$, $a_{2}$ is $|1,1,1,1\rangle$ where
$i,j=H$ or $V$. The other modes are initially in the vacuum state
$|0,0,0,0\rangle$. We propagate these modes through the system
and obtain the following expressions for the output
modes
\begin{eqnarray}
c_{Ho} & = & c_{H} \; \; \; \; \; \; \; \; \;
c_{Vo}  =  {{1}\over{\sqrt{2}}}(d_{1}'+d_{2}') \nonumber\\
t_{Ho} & = & {{1}\over{\sqrt{2}}}(t''+t''')  \; \; \; \; \; \;
t_{Vo}  =  {{1}\over{\sqrt{2}}}(t''-t''') \nonumber\\
a_{1o} & = & \sqrt{\eta_{3}} a_{1}''+\sqrt{1-\eta_{3}} v_{1}'
 \; \; \;
a_{2o}  =  \sqrt{\eta_{3}} a_{2}''+\sqrt{1-\eta_{3}} v_{2}'
\label{cnot}
\end{eqnarray}
where
\begin{eqnarray}
t'' & = & {{1}\over{\sqrt{2}}}(d_{1}'-d_{2}') \; \; \; \; \; \; \; \; \;
t'''  =  {{1}\over{\sqrt{2}}}(t_{H}-t_{V}) \nonumber\\
t' & = & {{1}\over{\sqrt{2}}}(t_{H}+t_{V}) \; \; \; \; \; \; \; \; \;
a_{1}''  =  \sqrt{\eta_{2}} a_{1}'+\sqrt{1-\eta_{2}} d_{1} \nonumber\\
a_{2}'' & = & \sqrt{\eta_{2}} a_{2}'+\sqrt{1-\eta_{2}} d_{2} \; \; \;
\; \; \; \; \; \;
a_{1}'  =  \sqrt{\eta_{1}} a_{1}+\sqrt{1-\eta_{1}} v_{1} \nonumber\\
a_{2}' & = & \sqrt{\eta_{1}} a_{2}+\sqrt{1-\eta_{1}} v_{2} \; \; \;
\; \; \; \; \; \;
v_{1}'  =  \sqrt{1-\eta_{1}} a_{1}-\sqrt{\eta_{1}} v_{1} \nonumber\\
v_{2}' & = & \sqrt{1-\eta_{1}} a_{2}-\sqrt{\eta_{1}} v_{2} \; \; \; \; \; \;
\; \; \;
d_{1}'  =  \sqrt{1-\eta_{2}} a_{1}'-\sqrt{\eta_{2}} d_{1} \nonumber\\
d_{2}' & = & \sqrt{1-\eta_{2}} a_{2}'-\sqrt{\eta_{2}} d_{2} \; \; \; \; \; \;
\; \; \;
d_{1}  =  {{1}\over{\sqrt{2}}}(c_{V}+t') \nonumber\\
d_{2} & = & {{1}\over{\sqrt{2}}}(c_{V}-t')
\label{cond1}
\end{eqnarray}
The logical statements of Eq.\ref{logic} can then be realized through
measurements of 4-fold coincidences. Thus if the initial state is
$|H\rangle_{c} |H\rangle_{t}$ then we find
\begin{eqnarray}
    \langle c_{Ho}^{\dagger} c_{Ho} \; \; t_{Ho}^{\dagger} t_{Ho} \; \;
    a_{1o}^{\dagger} a_{1o} \; \;
    a_{2o}^{\dagger} a_{2o}\rangle & = & {{1}\over{16}} \nonumber\\
    \langle c_{Ho}^{\dagger} c_{Ho} \; \; t_{Vo}^{\dagger} t_{Vo} \; \;
    a_{1o}^{\dagger} a_{1o} \; \;
    a_{2o}^{\dagger} a_{2o}\rangle & = & 0 \nonumber\\
     \langle c_{Vo}^{\dagger} c_{Vo} \; \; t_{Vo}^{\dagger} t_{Vo} \; \;
     a_{1o}^{\dagger} a_{1o} \; \;
    a_{2o}^{\dagger} a_{2o}\rangle & = & 0 \nonumber\\
     \langle c_{Vo}^{\dagger} c_{Vo} \; \; t_{Ho}^{\dagger} t_{Ho} \; \;
     a_{1o}^{\dagger} a_{1o} \; \;
    a_{2o}^{\dagger} a_{2o}\rangle & = & 0
    \label{cond2}
\end{eqnarray}
and similarly for the initial state $|H\rangle_{c} |V\rangle_{t}$ we find
\begin{eqnarray}
  \langle c_{Ho}^{\dagger} c_{Ho} \; \; t_{Vo}^{\dagger} t_{Vo} \; \;
    a_{1o}^{\dagger} a_{1o} \; \;
    a_{2o}^{\dagger} a_{2o}\rangle & = & {{1}\over{16}}
    \label{cond3}
\end{eqnarray}
with all other moments zero. However for initial state the
$|V\rangle_{c} |H\rangle_{t}$ we find
\begin{eqnarray}
     \langle c_{Vo}^{\dagger} c_{Vo} \; \; t_{Vo}^{\dagger} t_{Vo} \; \;
    a_{1o}^{\dagger} a_{1o} \; \;
    a_{2o}^{\dagger} a_{2o}\rangle & = & {{1}\over{16}}
    \label{cond4}
\end{eqnarray}
with the other moments zero and for the initial state
$|V\rangle_{c} |V\rangle_{t}$ we find
\begin{eqnarray}
     \langle c_{Vo}^{\dagger} c_{Vo} \; \; t_{Ho}^{\dagger} t_{Ho} \; \;
     a_{1o}^{\dagger} a_{1o} \; \;
    a_{2o}^{\dagger} a_{2o}\rangle & = & {{1}\over{16}}
\end{eqnarray}
with the other moments zero.
As expected the factor $1/16$ appears as we have employed two
NS gates each of which works on average 25\% of the time. It can also
be verified that injection of the control qubit in the superposition
states $(1/\sqrt{2})(|H\rangle\pm|V\rangle)$
with the target in $|H\rangle$ or $|V\rangle$ produces correlations
corresponding to the
4 entangled Bell states, as expected from quantum CNOT operation.

\section{Simplified Gate Operation}

A major
experimental advantage to this set-up, as compared to the test circuit
suggested in
Ref. \cite{kni00}, is that we can work in the coincidence
basis. This allows low efficiency detectors and spontaneous single photon
sources to be used to demonstrate the basic operation of the gate. Of 
course incorporating 
these gates in a scalable system as
discussed in Section II requires one to know that the gate has 
successfully operated without destroying the output. 
It is straightforward to show
from Eqs.\ref{cnot} that detection of one and only one photon in
modes $a_{1o}$ and $a_{2o}$ and no photons in modes $v_{1o}$ and
$v_{2o}$ is sufficient to ensure successful operation of the gate
without disturbing the control and target outputs. However low-loss, 
0, 1, 2-photon discriminating detection would be needed to operate in 
this way.

Even in the coincidence basis the above implementation represents a major
technological challenge. Four nested interferometers must simultaneously be
mode matched and locked to sub-wavelength accuracy over the operation
time of the gate. A major
simplification is achieved by operating the NS gates in a biased
mode. The idea is to set the reflectivities $\eta_{1}$ and $\eta_{3}$
in the NS gates to one, i.e. totally reflective. This removes the
interferometers from both the NS gates, greatly reducing the
complexity of the gate. Summing over the paths as before we find that
the NS operation becomes
\begin{equation}
|\psi\rangle=\alpha |0\rangle+\beta |1\rangle+
\gamma |2\rangle \to |\psi'\rangle=
\sqrt{\eta_{2}} \alpha |0\rangle+(1-2 \eta_{2}) \beta |1\rangle-
\sqrt{\eta_{2}}(2-3 \eta_{2})\gamma |2\rangle
\label{NSS}
\end{equation}
when $\eta_{1}=\eta_{3}=1$. There is no solution such that the ``0'',
``1'' and ``2'' components scale equally, so the gate is biased. 
However this problem can be
solved by placing an additional beamsplitter in the beam path with a 
vacuum input and conditioning on no photons appearing at its output. 
Now we find
\begin{equation}
 |\psi'\rangle=
\sqrt{\eta_{2}} \alpha |0\rangle+\sqrt{\eta_{7}}(1-2 \eta_{2}) \beta |1\rangle-
\eta_{7} \sqrt{\eta_{2}}(2-3 \eta_{2})\gamma |2\rangle
\label{NSSs}
\end{equation}
where $\eta_{7}$ is the reflectivity of the additional beamsplitter. 
Remarkably the additional degree of freedom allows the gate to be rebalanced 
such that exact NS operation is achieved without an interferometric 
element. The trade-off is a small reduction in the probability of success.
Solving we find $\eta_{2}=(3-\sqrt{2})/7$ and $\eta_{7}=5-3 \sqrt{2}$ 
gives NS operation with a success probability of $\eta_{2}\approx 0.23$.

There is considerable flexibility in how the simplified gate is 
employed in the CNOT. 
One of a number of possible scenarios is shown in Fig.4. The NS gates 
of Fig.3 have been
replaced by the beamsplitters B5 and B6 which have reflectivities
$\eta_{2}$. Additional beamsplitters, B7 and B8, of reflectivities
$\eta_{7}$ have been inserted in beams $c_{V}$ and $t'$ respectively.
The state of the system at point $z$ in Fig.4 (conditional on a single
photon being detected at outputs $a_{1o}$ and $a_{2o}$ {\it and} no
photons appearing at outputs $v_{7o}$ and $v_{8o}$) is given by
\begin{equation}
|\psi \rangle_{y}={{1}\over{\sqrt{2}}}\eta_{2}|1001 \rangle \pm
\sqrt{\eta_{2} \eta_{7}}(1-2 \eta_{2}){{1}\over{2}}(|1100
\rangle - |1010 \rangle )
\end{equation}
if the control is initially in $|H \rangle$ and
\begin{equation}
|\psi \rangle_{y}={{1}\over{2}}(\sqrt{\eta_{2} \eta_{7}}(1-2 \eta_{2})
(|0101 \rangle +
|0011 \rangle) \mp (\eta_{7} \eta_{2}(2-3 \eta_{2})(|0200
\rangle - |0020 \rangle) ))
\end{equation}
if the control is initially in $|V \rangle$. Choosing as before 
$\eta_{2}=(3-\sqrt{2})/7$ and $\eta_{7}=5-3 \sqrt{2}$ we obtain CNOT
operation with a probability $\eta_{2}^{2} \approx 0.05$.
The operation of the gate can also still be described by Eq.\ref{cnot} but
with $\eta_{1}=\eta_{3}=1$ and the substitutions
\begin{eqnarray}
c_{V}  =
{{\sqrt{\eta_{7}}}\over{\sqrt{2}}}c_{V}'+\sqrt{1-\eta_{7}
} v_{8}, \; \; \; \;
t' =
{{\sqrt{\eta_{7}}}\over{\sqrt{2}}}(t_{H}+t_{V})+\sqrt{1-\eta_{7}
} v_{7}
\end{eqnarray}
where now $c_{V}'$ is the initial state of the control's vertical
polarization mode. All the conditional moments of
Eq.\ref{cond1}-\ref{cond4} are reproduced but with the probabilities
of the non-zero moments reduced from $1/16$ to approximately $1/20$.
All other properties of the original gate are retained.

\section{Conclusion}

The efficient linear optics computation scheme of Ref.\cite{kni00}
appears exciting in principle but daunting in practice. However we have shown
that by adopting a CNOT test architecture the basic principles of the
scheme can be tested with present technology. Four-photon
experiments with spontaneous sources are difficult, but have been
achieved \cite{zeil}. Basically such experiments utilize events where 
by chance two down converters simultaneously produce pairs.
The use of our simplified scheme would reduce the stability issues in
such an experiment significantly with only a small decrease in
probability of success. Calculations using Eq.\ref{cnot} 
show that operation is not 
critically dependent on experimental parameters. For example 2\% 
errors in beamsplitter ratios only lead to fractions of a percent 
errors in gate operation.

In the longer term the greater
simplicity of our gate is likely to play a significant
role in scalable architectures
when the required single photon sources and detectors become available.

This research was supported by the Australian Research Council.

\begin{figure}
    
\caption{Review of Teleportation of Gates. (a) shows a basic 
teleportation circuit. (b) shows two such circuits with a CNOT 
implemented post teleportation. (c) shows the effect of commuting the 
CNOT through the $\sigma_{x}$ ($X$) and $\sigma_{z}$ ($Z$) operations. 
The dotted 
line encloses the entanglement resource which could be produced using 
non-deterministic gates.}
\end{figure}

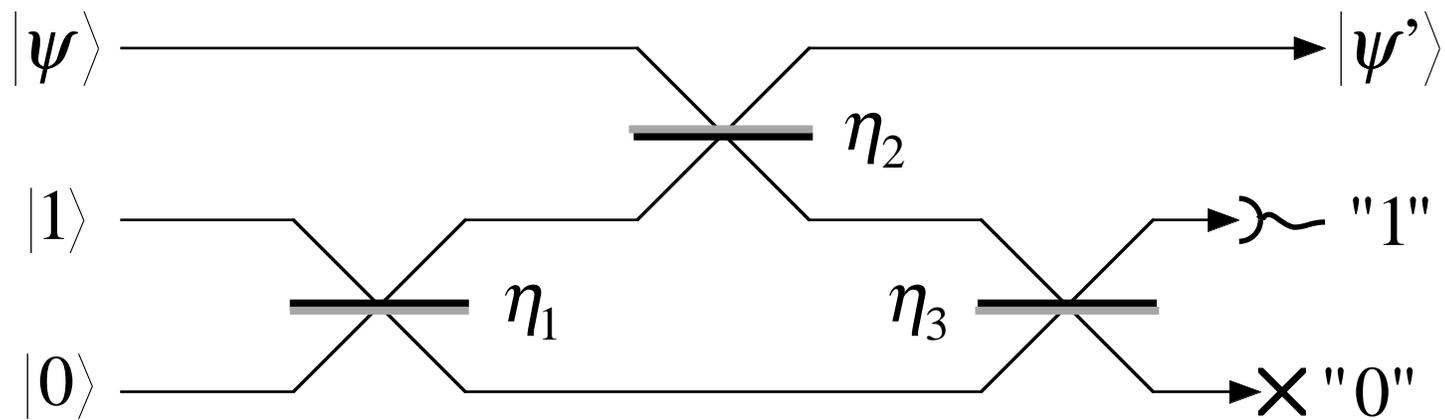
\begin{figure}

\caption{Schematic of NS gate. Grey indicates the surface from
which a sign change occurs upon reflection. The use of this 
beamsplitter phase convention is convenient but not essential.}
\end{figure}

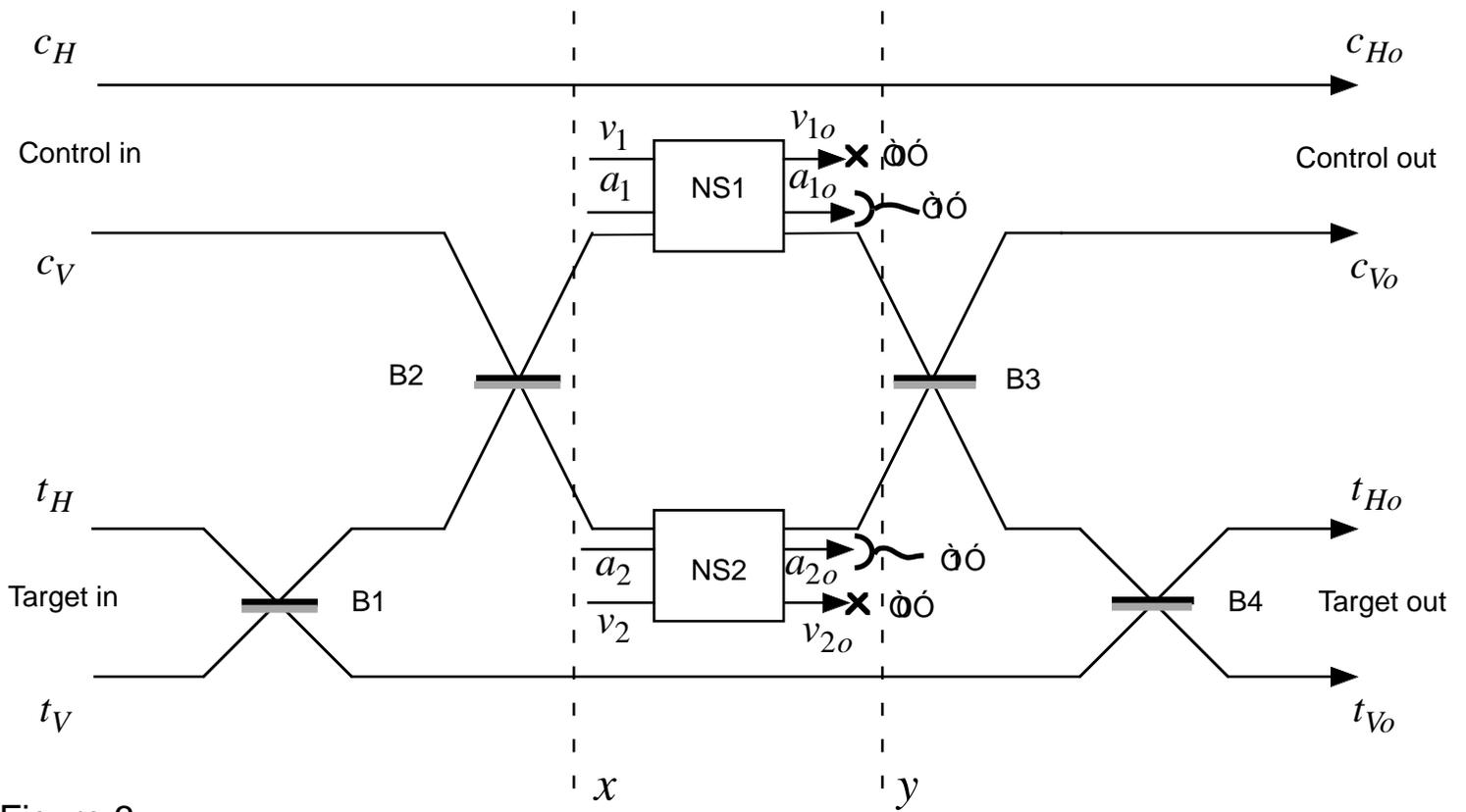
\begin{figure}

\caption{Schematic of CNOT gate. Grey indicates the surface from
which a sign change occurs upon reflection. Note that if B1 and B4 
were not present the gate would implement a control sign shift. B1 and 
B4 paly the role of Hadamard gates coverting sign shift to CNOT operation}
\end{figure}

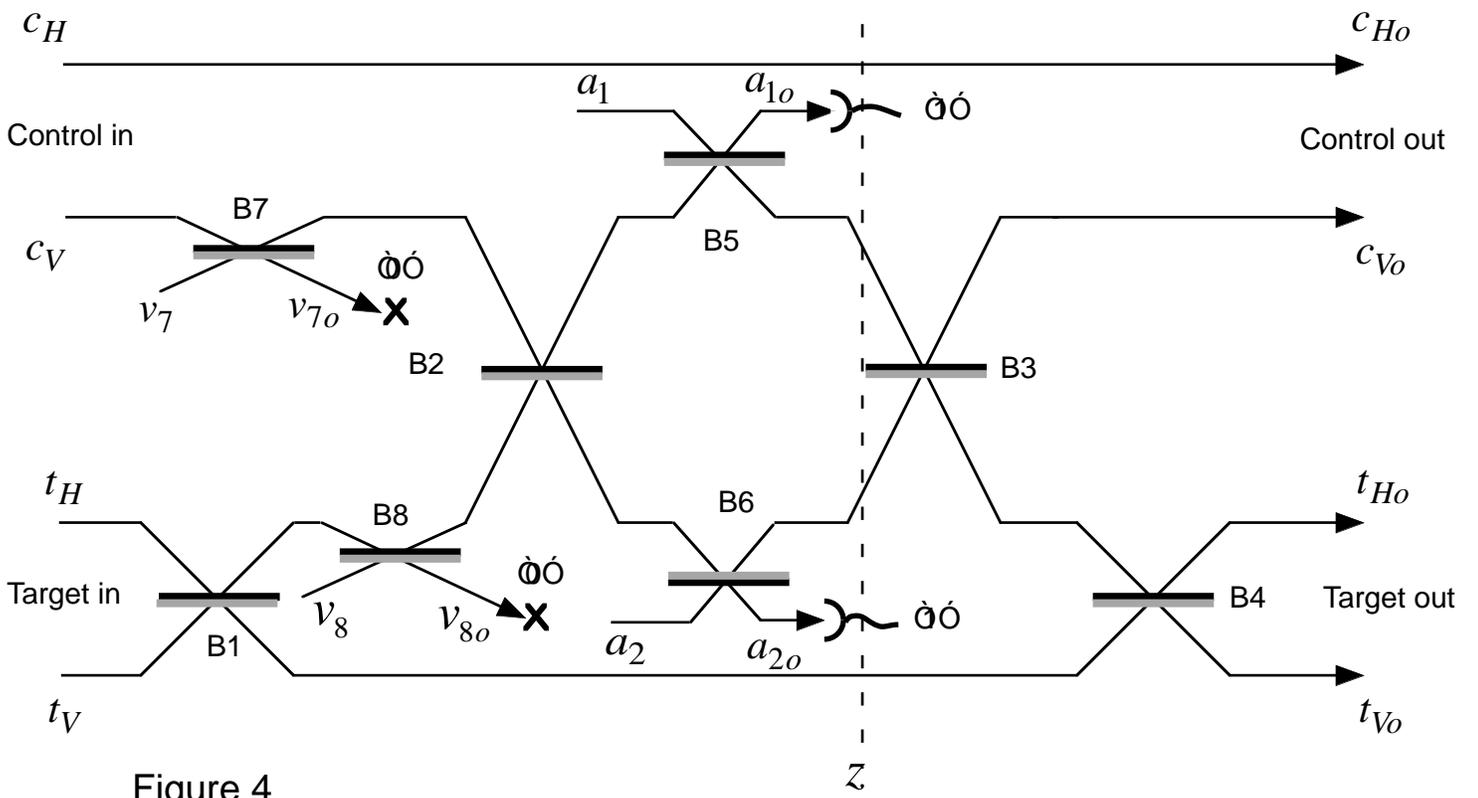
\begin{figure}

\caption{Schematic of simplified CNOT gate. Grey indicates the surface from
which a sign change occurs upon reflection.}
\end{figure}

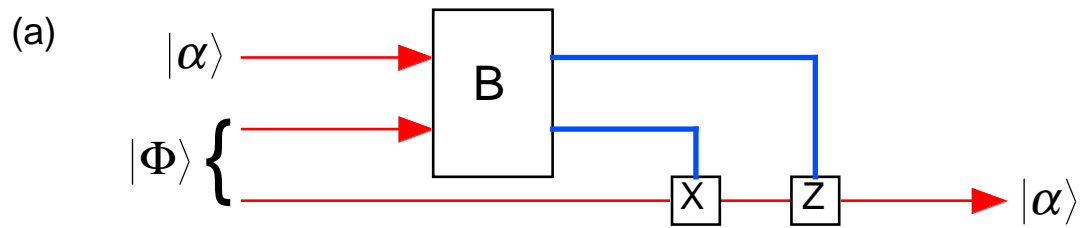
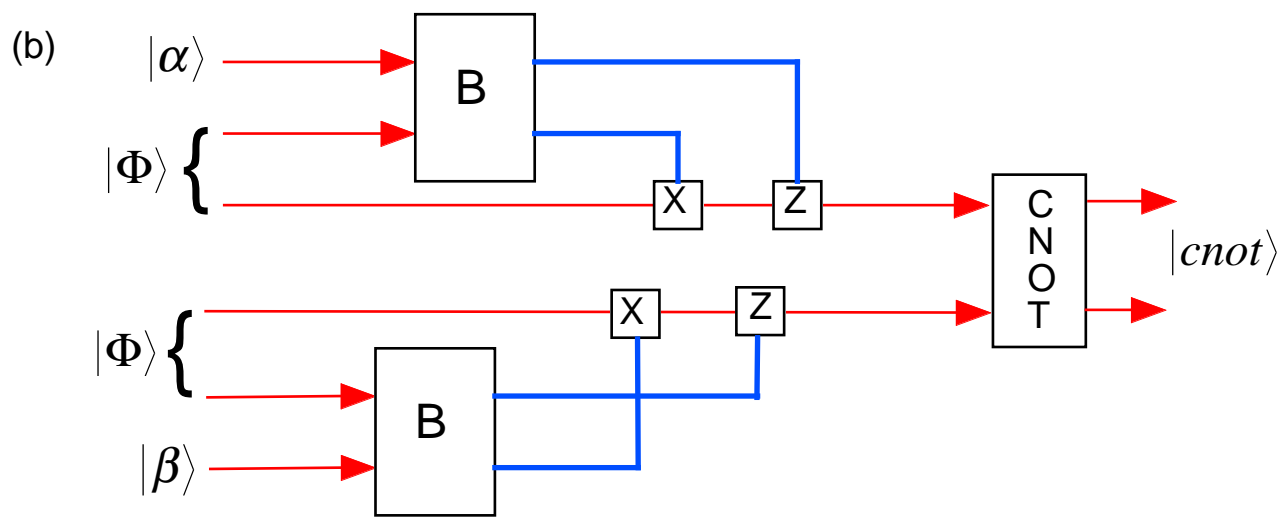
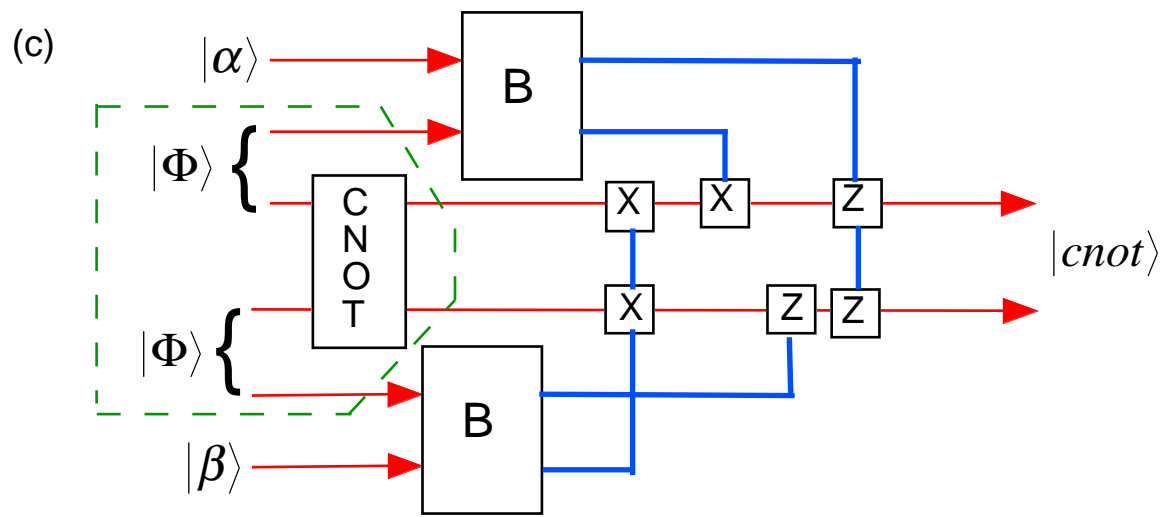
\end{document}